# Mapeamento de Intensidade da Pulsação Sanguínea por Vídeo
## Blood Pulsation Intensity Video Mapping

Pedro Henrique de Marco Borges[1], George C. Cardoso[1]

[1]Departamento de Física, Universidade de São Paulo, Ribeirão Preto, Brasil

**Resumo**
Neste estudo medimos a frequência cardíaca de forma não invasiva, remota e passiva e determinamos o mapa da atividade de pulsação sanguínea numa região de interesse (ROI) da pele. A ROI utilizada foi o antebraço de um voluntário. O método utiliza uma câmara de vídeo comum e luz visível como processo de aquisição do vídeo durante menos de 1 minuto. A frequência cardíaca média encontrada tem erro em torno de 1 batimento por minuto em relação a uma medida simultânea usando fotopletismografia de lóbulo de orelha. A partir dos sinais extraídos das imagens de vídeo, determinamos um mapa da intensidade da pulsação sanguínea na superfície da pele. Apresentamos detalhes experimentais e de processamento dos sinais, assim como limitações da técnica.

**Palavras-chave**: frequência cardíaca; mapeamento da circulação sanguínea; perfusão sanguínea; processamento de vídeo.

*Abstract*
*In this study, we make non-invasive, remote, passive measurements of the heart beat frequency and determine the map of blood pulsation intensity in a region of interest (ROI) of skin. The ROI used was the forearm of a volunteer. The method employs a regular video camera and visible light, and the video acquisition takes less than 1 minute. The mean cardiac frequency found in our volunteer was within 1 bpm of the ground-truth value simultaneously obtained via earlobe plethysmography. Using the signals extracted from the video images, we have determined an intensity map for the blood pulsation at the surface of the skin. In this paper we present the experimental and data processing details of the work and well as limitations of the technique.*

## 1. Introdução

O monitoramento de sinais vitais nas áreas clínica e hospitalar é de extrema importância. Atualmente, esses monitoramentos são feitos com base em sensores que envolvem contato físico, tais como em eletrocardiografia (ECG). Porém esses sensores podem causar danos à pele de recém-nascidos prematuros e em pacientes que sofreram queimaduras devido à sensibilidade da pele, inviabilizando o uso dessa técnica para alguns tipos de diagnósticos clínico[1]. Dessa forma, a pesquisa em técnicas não invasivas tem se intensificado nos últimos anos. Uma das técnicas mais ativamente pesquisadas na atualidade para medida remota de pulsação cardíaca é o uso de vídeos produzidos por câmeras no espectro de luz visível[2,3,4]. Esse tipo de monitoramento evita os possíveis danos à pele e infecções relacionadas.

O princípio clínico que permite o uso de câmeras de vídeo para o monitoramento de pulsação cardíaca é a fotoplestimografia (PPG) óptica. A PPG consiste numa técnica óptica para realizar medições de diferença de volume sanguíneo durante a pulsação sanguínea. O efeito é o mesmo utilizado para determinar a pulsação em oxímetros de dedo (PPG de dedo). À medida que o coração bombeia o sangue, o volume sanguíneo nas artérias e capilares muda por uma pequena quantidade em sincronia com o ciclo cardíaco. Até recentemente entendia-se que o mecanismo de funcionamento dos PPG de dedo era por variação na absorção ou no espalhamento óptico devido ao aumento periódico do volume sanguíneo nos vasos[5-8]. Mais recentemente foi mostrado que a mudança no volume sanguíneo resulta num pequeno aumento na pressão exercida sobre os tecidos locais, causando sutil variação de cor (croma) na pele[9]. Isso permite que pulsações sanguíneas, a profundidades maiores que permitidas pela penetração de luz visível, possam gerar mudanças das propriedades de reflexão da luz na superfície da pele. Na PPG por vídeo, a variação da croma em um ciclo cardíaco – variação que não é observável a olho nu – é detectada na imagem processada do vídeo. A sensibilidade cromática de câmeras pode ser muito superior à sensibilidade da visão humana, mesmo para as câmeras RGB de 8 bits por canal. Assim, o objetivo do uso de câmeras é adquirir imagens de vídeo que conterão sinal temporal referente à pulsação local. Uma das aplicações que tem sido discutidas para o uso de câmeras é um mapeamento de intensidade da pulsação sanguínea para a caracterização de possíveis micro lesões na vascularização[10,13]. Essa aplicação já é bem estabelecida nas técnicas de Doppler, tanto com ultrassom[11] quanto com laser[12], para análise de perfusão sanguínea. A utilização de vídeo representa uma tecnologia alternativa de fácil uso, menos invasiva e de mais baixo custo[14,16].

Neste trabalho apresentamos um método para determinação da frequência cardíaca e um novo algoritmo para mapeamento da perfusão sanguínea através de imagens de vídeo da superfície externa do corpo humano. Diferente de trabalho de trabalhos anteriores (ver, por exemplo *Kamshilin et al. (2015)*[9]) não utilizamos iluminação especial, polarizadores cruzados, nem imagens de vídeo sem compressão *(raw)*.

## 2. Materiais e Métodos

Para a realização das filmagens fez-se uso de uma câmera Canon 50D adaptada para aquisição de vídeo com o firmware *Magic Lantern*, e operando a 30 quadros por segundo (30 fps), com cartão de memória *compactflash* 1000x, 16 GB. A câmara foi suportada por um tripé. Uma lâmpada incandescente (General Electric, *Reveal Flood* refletora de 100 watts) foi utilizada como fonte de iluminação. Uma cadeira reclinada com braço para apoio dos braços do voluntário foi utilizada para minimizar movimentos indesejados. Para o processamento do vídeo fez-se uso de um computador com processador Intel Core i7, e software MATLAB (versão 2010, com pacote de processamento de imagens). Para comparação dos resultados obtidos foi usado um sensor óptico de lóbulo de orelha para PPG (HeartMath Institute, 30 pinos para iOS) para aquisição contínua do sinal de PPG simultaneamente à aquisição das imagens.

Para aquisição das imagens, inicialmente o voluntário senta-se reclinado, tentando ficar o mais confortável possível na cadeira, para evitar movimentos que levem a causar sinais espúrios nas imagens, o procedimento é análogo ao adotado por outros pesquisadores nesta área[1-10,18].

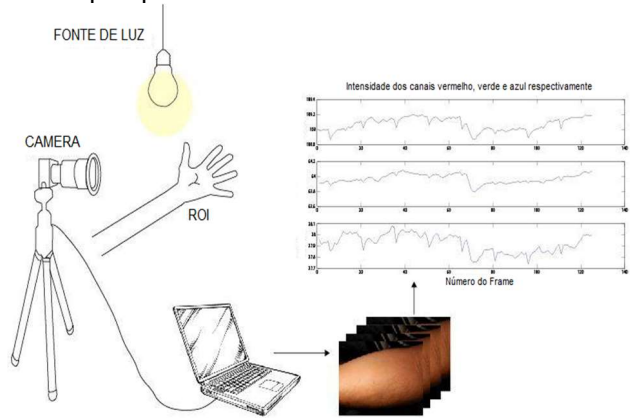

**Figura 1:** Procedimento experimental

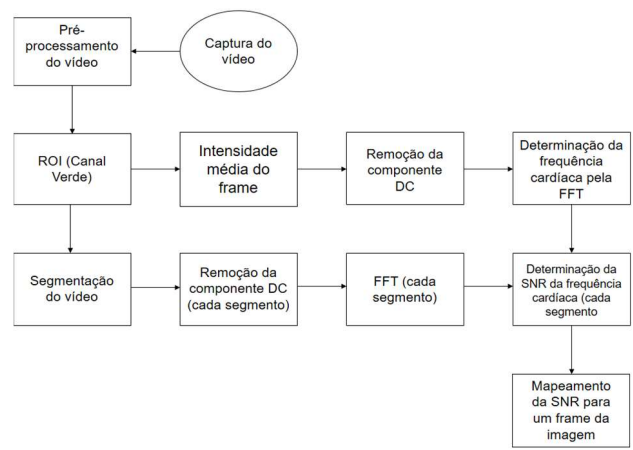

**Figura 2:** Fluxograma do processamento do vídeo

A fonte de iluminação é pré-aquecida por pelo menos 5 minutos para evitar flutuação na intensidade. Aguardamos a estabilização dos batimentos cardíacos do voluntário, observando através do sinal do PPG de lóbulo de orelha. Essa estabilização visa minimizar flutuação da frequência cardíaca no intervalo de aquisição das imagens a fim de simplificar a análise de sinal neste projeto. Em seguida focalizamos a câmera na região de interesse (ROI) com uma lente de distância focal de cerca entre 30 mm e 40 mm, mantida a uma distância de aproximadamente 25 cm do voluntário. Essa distância é a que proporcionou um melhor sinal/ruído com a iluminação utilizada. Uma vez que usamos luz incandescente, aumentar a iluminação poderia causar algum desconforto no voluntário devido a um pequeno aquecimento da pele. Vários vídeos de 1 minuto de duração em condições ligeiramente diferentes foram obtidos para análise. Esse tempo é compatível com o tempo utilizado por outros autores na área[19-20]. A Figura 1 mostra esquematicamente uma representação do sistema de aquisição e processamento das imagens de vídeo.

Após transferência dos arquivos de vídeo para o PC, o processamento foi feito em MATLAB. A Figura 2 mostra um fluxograma simplificado do algoritmo utilizado. Após seleção da região de interesse, o canal verde do vídeo é selecionado, pois proporciona melhor relação sinal/ruído que os canais vermelho e azul. A intensidade média de cada frame é calculada, o que gera uma série temporal de intensidades oscilando em fase com os batimentos cardíaco, conforme Figura 3. Um filtro Butterworth passa baixa de primeira ordem (com atenuação de -3dB em 23 bpm), foi utilizado para obter a componente de variação lenta do sinal – tais flutuações lentas de intensidade são oriundas de pequenos movimentos, da respiração e de variação na luminosidade da sala devido a movimentos do experimentalista. A frequência de corte do filtro foi encontrada empiricamente *a posteriori* e é adequada para uma faixa larga de frequências cardíacas. A componente de variação lenta é subtraída do sinal original, gerando o sinal mais uniforme e centrado no zero, como mostrado na Figura 3 (meio). Os primeiros 200 pontos (ou cerca de 6,7 segundos de vídeo) foram desprezados pois estão na região onde o filtro Butterworth ainda não convergiu. Após essas modificações é aplicada uma transformada discreta de Fourier (FFT) para encontrar o valor da frequência central dos batimentos cardíacos na ROI como um todo. Em nosso experimento, a fim de validar os resultados, o valor da pulsação é verificado contra os resultados obtidos pelo fotopletismógrafo de lóbulo de orelha.

Para construir o mapa espacial de intensidade da pulsação sanguínea, o vídeo é segmentado em m x n partições como mostrado na Figura 4. Para isso, criou-se uma matriz auxiliar (n x m), em que n e m são os números de partições desejadas nas direções horizontal e vertical. Para determinar a intensidade relativa do sinal dos batimentos cardíacos em cada partição, é calculada para cada partição a área do pico da FFT (cuja frequência foi determinada antes da segmentação, como no procedimento do parágrafo acima). A área do pico é então dividida pela área do espectro inteiro de 0 a 200 bpm, conforme Eq. (1):

$$SNR_{m,n} = \frac{\int_{BPM-1}^{BPM+1} I_{m,n}(f)df}{\int_{0}^{200} I_{m,n}(f)df}, \quad (1)$$

onde $I_{m,n}(f)$ é o módulo da FFT como função da frequência cardíaca. BPM±1 representa a largura a meia altura de |Y(f)| (Figura 3). A Figura 4 mostra um exemplo de matriz dos valores de $SNR_{m,n}$, ou seja, das intensidades relativas do sinal da pulsação sanguínea na região observada.

O tempo total de processamento, no programa em MATLAB escrito sem otimização foi de cerca de 50 segundos em laptop de 16 GB de RAM, processador Intel i7.

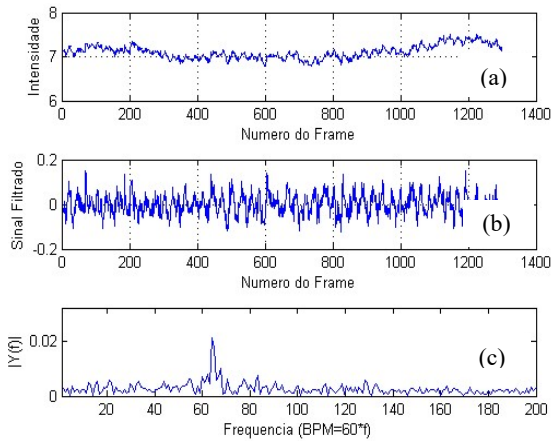

**Figura 3:** (a) Intensidade média do canal verde na ROI. (b) Sinal com variação lenta subtraída. (c) Módulo da FFT de (b).

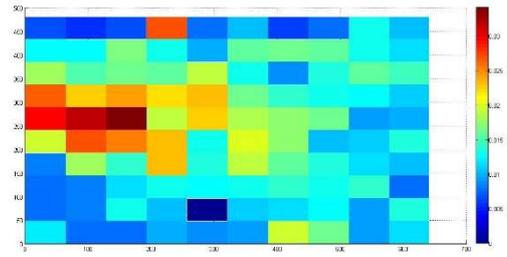

**Figura 4:** Mapa com as amplitudes relativas da pulsação sanguínea na região de interesse.

O pesquisador principal deste trabalho (GCC) serviu como o voluntário neste trabalho. Não há riscos envolvidos na aquisição do vídeo, pois se trata de técnica não-invasiva, remota e passiva. Portanto, o trabalho está de acordo com os princípios éticos da Declaração de Helsinki[15].

### 3. Resultados e Discussão

O resultado da pulsação cardíaca das imagens correspondentes à Figura 3 é de 65 ±1 BPM. O valor concorda com o obtido através do sistema pletismográfico de lóbulo de orelha 64 ± 2 BPM (frequência cardíaca média, máxima e mínima no intervalo de tempo do vídeo). Esse resultado é comparável ao obtido por Yu *et al.* (2015)[16]. Entretanto, em alguns dos testes realizados, a FFT da área de interesse (antebraço inteiro) não apresentou o pico dos batimentos cardíacos. Verificou-se que o sinal é proporcional a intensidade da iluminação utilizada, porém a reflexão especular na pele tende a saturar a câmera, eliminando o sinal dos batimentos cardíacos. O ângulo entre a câmera e a fonte de luz em relação a superfície da pele é importante (na ausência de polarizadores cruzados). Testamos também lâmpadas de LED "brancas" sem obter sinais dos batimentos cardíacos. Diante disso, recomendamos o uso de LEDs monocromáticos para obter boa iluminação apenas nos comprimentos de onda de interesse, para minimizar a potência total dissipada e saturação da câmera em comprimentos de onda indesejados. LEDs verdes foram utilizados por *Kamshilin et al.[9],* porém, como discutiremos adiante, o uso de três cores de LEDs de espectros adequados melhoraria o sinal. Não encontramos na literatura uma análise da dependência temporal do espectro visível de reflexão da pele em função da pulsação sanguínea. Tal estudo seria útil para especificar os três comprimentos de onda ótimos para uso na iluminação da pele e que correspondessem aos canais vermelho, verde e azul (R, G, B) do vídeo, que maximizasse o sinal da fotopletismografia.

Resultados de um mapa de intensidade da pulsação é mostrado na Figura 5. No lado esquerdo da figura mostramos um mapa de intensidade da pulsação superposto à imagem do antebraço. A fim de deixar a imagem menos carregada, mostramos as cores falsas somente nas áreas onde o sinal da pulsação está acima de um certo limiar de amplitude. Na Figura 5 a cor mais escura indica um sinal de pulsação menor. A região branca na parte inferior da Figura 5 (esquerda) é um artefato pois é um quadro usado para determinar a orientação da imagem. Por outro lado, a região branca na parte superior da mesma figura, indicando um sinal de pulsação, foi identificado no vídeo como um objeto sobre a perna do voluntário que está pulsando em fase com a pulsação sanguínea. Isso ocorre porque em alguns casos não é possível distinguir uma vibração de uma mudança cromática sem um processamento cuidadoso. O mapa de intensidade do sinal de pulsação apresentado não está normalizado pela luminância local, não representando nesse caso o mapa da atividade de pulsação sanguínea diretamente. As Figuras 4 e 5 não correspondem ao mesmo experimento.

Pode-se observar na Figura 3, o sinal dos batimentos cardíacos na imagem de vídeo é bastante ruidoso. Isso está em contraste com os resultados de Kumar et al. (2015)[1] ou Kamshilin et al. (2015)[9]. Por outro lado, mostramos que é possível obter os sinais usando luz não monocromática (incandescente), não utilizamos polarizadores cruzados (que evitaria reflexão que não contém o sinal de interesse) e principalmente utilizamos imagens de vídeo comprimidas (Canon MOV tipo H.264). Para popularizar vídeo pletismografia, o ideal é obter técnicas capazes de aceitar vídeos comprimidos, pois somente câmeras especializadas e de alto custo conseguem produzir vídeos não comprimidos (raw) e de alta resolução, além do custo de memória e de tempo de processamento associados aos vídeos raw. O desafio principal dos vídeos comprimidos é a compressão na cor – os canais de cor com o protocolo H.264 não atualizam quadro a quadro.

Uma forma de melhorar a relação sinal ruído com as imagens obtidas é aplicar a técnica de separação cega (BSS) de fontes ou análise das componentes principais[16,23]. Observamos que os canais azul e vermelho das imagens de vídeo também contêm o sinal da pulsação cardíaca. O uso de BSS poderá extrair o máximo do sinal da pulsação usando as três componentes do vídeo. Um sinal temporal menos ruidoso permitiria a determinação do equivalente aos intervalos RR. Assim seria possível aplicar o método que apresentamos mesmo quando a frequência cardíaca mude muito rapidamente – um caso que evitamos durante a aquisição dos dados.

No prosseguimento deste trabalho, implementaremos as melhorias acima discutidas assim como utilizaremos uma câmera com velocidade de pelo menos 100 quadros por segundo para resolver o tempo de chegada do pulso sanguíneo nas várias partições da região da área de interesse.

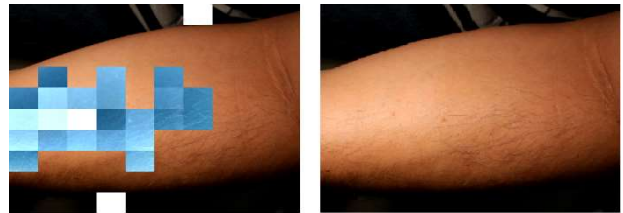

**Figura 5**: À esquerda: Mapa da intensidade do sinal da pulsação sanguínea superposta na imagem real. À direita: a imagem física.

É conhecido que existe ambiguidade na fase da pulsação das várias regiões[24]. Porém, numa extensão do mapeamento que aqui desenvolvemos, uma correlação cruzada dos sinais temporais das várias partições da ROI poderá fornecer um mapa do tempo de chegada do pulso, mapeando a propagação do pulso sanguíneo. Tal mapa potencialmente permitiria observar anomalias na perfusão sanguínea. Essas anomalias poderiam estar relacionadas a enfermidades do sistema circulatório local ou nos tecidos em torno da região.

## 5. Conclusões

Este trabalho adiciona à literatura um novo método capaz de fornecer um mapa da pulsação sanguínea numa região da pele. Discutimos os desafios e possíveis melhorias tanto na instrumentação utilizada quanto no método de processamento das imagens. Com a continuação das pesquisas, acreditamos que num futuro próximo câmeras, mesmo de baixo custo, poderão ser lugar-comum para auxílio em diagnósticos clínicos e ambulatoriais.

**Contato:**
George C. Cardoso
Departamento de Física -FFCLRP
Universidade de São Paulo
Av. Bandeirantes, 3900
Monte Alegre, Ribeirão Preto – SP
14040-901 Brasil
*E-mail: gcc@usp.br*